\documentclass[twocolumn,floatfix,prb,showpacs,preprintnumbers, footinbib]{revtex4}
\usepackage{amssymb}
\usepackage[dvips]{graphicx}
\usepackage{dcolumn}
\usepackage{bm}

\begin{document}

\title{Two-phonon assisted exciton spin relaxation due to exchange interaction in spherical quantum dots}
\author{P. Nah\'alkov\'a$^{1,2}$, D. Sprinzl$^1$, P. Mal\'y$^1$, P. N\v{e}mec$^1$, V. N. Gladilin$^2$, J. T. Devreese$^2$}
\affiliation{$^1$Faculty of Mathematics and Physics, Charles University in Prague, Ke Karlovu 3, 121 16 Prague 2, Czech Republic\\
$^2$Theoretische Fysica van de Vaste Stoffen, Departement Fysica, Universiteit Antwerpen, Groenenborgerlaan 171, B-2020 Antwerpen, Belgium}

\begin{abstract} Spin relaxation in undoped quasi-spherical CdS quantum dots at zero magnetic fields is investigated using time- and polarization-resolved transient absorption measurements. Unlike in previous studies of these systems, the measured signals were corrected for spin-insensitive contributions to the exciton bleaching dynamics, {allowing us to determine} the pure spin-related exciton dynamics. To explain the observed room-temperature spin-relaxation time of several nanoseconds, we propose a novel mechanism based on intralevel exciton transitions with the emission of one LO phonon, the absorption of another LO phonon, and an electron spin flip, which is driven by the electron-hole exchange interaction. The transition rates, calculated in the present work for different sizes of quantum dots and temperatures, are in fair agreement with our experimental results.

\end{abstract}

\date{\today}
\maketitle


Spin relaxation and decoherence in quantum dots (QDs) have attracted increasing attention over the last few years. This is i.a. due to the long spin-decoherence times observed in QDs, which have led to suggestions about their possible applications in quantum computation \cite{Loss98,Calarco03}. The underlying mechanisms of spin relaxation in semiconductor QDs are still being debated (see Ref.~3 
for a recent review). The term \textquotedblleft quantum dot\textquotedblright\ is actually used for a variety of different objects. While most device concepts assume highly symmetric QDs, real QDs are usually strongly anisotropic. For example, the most often investigated self-assembled QDs have a base elongated along the [110] axis \cite{Astakhov06}, and QDs formed by interface fluctuations in narrow quantum wells are elongated along the [$\bar{1}$10] axis \cite{Gammon96}. As a result, the anisotropic exchange interaction splits the $|\pm 1\rangle $ radiative exciton doublet, with opposite total angular momentum projections, into two linearly polarized eigenstates $|X\rangle =\left( |1\rangle +|-1\rangle \right) /\sqrt{2}$ and $|Y\rangle=\left( |1\rangle -|-1\rangle \right) /(i\sqrt{2})$. On the other hand, chemically synthesized QDs~\cite{Gupta01,Gupta02,Stern05,Ekimov96} are nearly spherical, retaining higher symmetry, and therefore the exciton states $|\pm 1\rangle $ are degenerate eigenstates of the Hamiltonian. Research on chemically synthesized quasi-spherical QDs is motivated also by their possible use as building blocks for constructing artificial solids of semiconductor QDs in the bottom-up self-assembly approach \cite{Ouyang03}. However, there exist only few reports on spin relaxation in quasi-spherical QDs. Measurements of differential transmission at zero magnetic field in neutral QDs revealed rather small circular polarization of the signal, from which only qualitative conclusions could be drawn \cite{Gupta01,Kulakovskii03}. The results retrieved from the decay of the Faraday rotation indicate that the spin dynamics is considerably slower in quasi-spherical QDs ~\cite{Gupta01,Gupta02,Stern05} than in anisotropic QDs~\cite{Tartakovskii04,Gotoh03,Stievater02}. However, as the signal of Faraday rotation contains also contributions from the spin-insensitive exciton dynamics, these experiments cannot be used for the precise determination of the electron spin-relaxation time $T_{1}$ in quasi-spherical QDs. The majority of published works about spin relaxation in QDs concentrate on low-temperature measurements, where most of the spin-relaxation processes are strongly suppressed~\cite{Paillard01}. In the present paper, we report on spin relaxation (corrected for a finite carrier life time) for neutral excitons in quasi-spherical CdS QDs with the focus on room temperature, which is most appealing for potential applications. We observed two quite distinctive [(sub)picosecond and nanosecond] components in the measured dynamics of the circular polarization of the differential transmission signal, which are attributed in our model to relaxation processes without and with electron spin flip, respectively. {The dominant} mechanism of the dynamics, observed at the nanosecond time scale, is identified as caused by two-LO-phonon intralevel transitions with electron spin flip, driven by the electron-hole exchange interaction.

\textit{Experiment.} The samples used in our experiments were CdS QDs in a glass matrix, a well-known model material of mutually isolated quasi-spherical QDs \cite{Ekimov96}. Here we report results for glass filters made by Hoya and Schott, which contain QDs with wurtzite lattice, volume filling factor of about 0.1\% and typical radii 1.9~nm (sample GG435), 2.1~nm (sample Y-44), and 2.4~nm (sample Y-46). Our findings are quite general for this type of material as was verified by our experiments with samples made by different manufacturers (Corning, Toshiba). The distribution of the QD sizes (as determined from TEM analysis) can be described well by the lognormal function (e.g., for sample Y-44 we obtained a distribution with average radius 2.1~nm and standard deviation 0.27). The spin dynamics of the carriers were measured using the time- and polarization-resolved transient absorption technique of ultrafast laser spectroscopy. Spin polarization of optically excited carriers was achieved by absorption of circularly polarized pump pulses and the carrier dynamics was deduced from the measured dependence of the probe-pulse transmission on the time delay between pump and probe pulses. As a measure of transmission changes we used the differential transmission $\Delta T/T_{0}=(T_{E}-T_{0})/T_{0}$, where $T_{E}$ ($T_{0}$) is the transmission with (without) the pump pulse. A Ti-sapphire laser with 82~MHz repetition rate was the source of 80~fs pulses. The pump power was kept at a low level, where the dynamics is independent of the pump intensity (on average much less than one electron-hole pair per QD and per laser pulse is photoexcited). The nonequilibrium spin polarization is reflected in the difference between the differential transmission dynamics measured with probe pulses co- and counter-polarized with respect to the pump pulses (called \textit{difference} signal in the following). This \textit{difference} signal can be directly measured using a lock-in amplifier and a photo-elastic modulator, which periodically modulates the helicity of the pump pulses. However, the \textit{difference} signal decays not only due to carrier spin flips but also because of energy relaxation and recombination of carriers. On the other hand, the sum of the differential transmission dynamics measured with co- and counter-circularly polarized probe pulses (\textit{sum} signal) is sensitive only to carrier recombination and energy relaxation. Therefore, the ratio of the \textit{difference} and \textit{sum} signals, which {is called here circular polarization of the signal ($P_{C}$)}, can be used to determine the spin-relaxation time $T_{1}$.

\begin{figure}[tbp]
\centering \includegraphics[width=8.2cm]{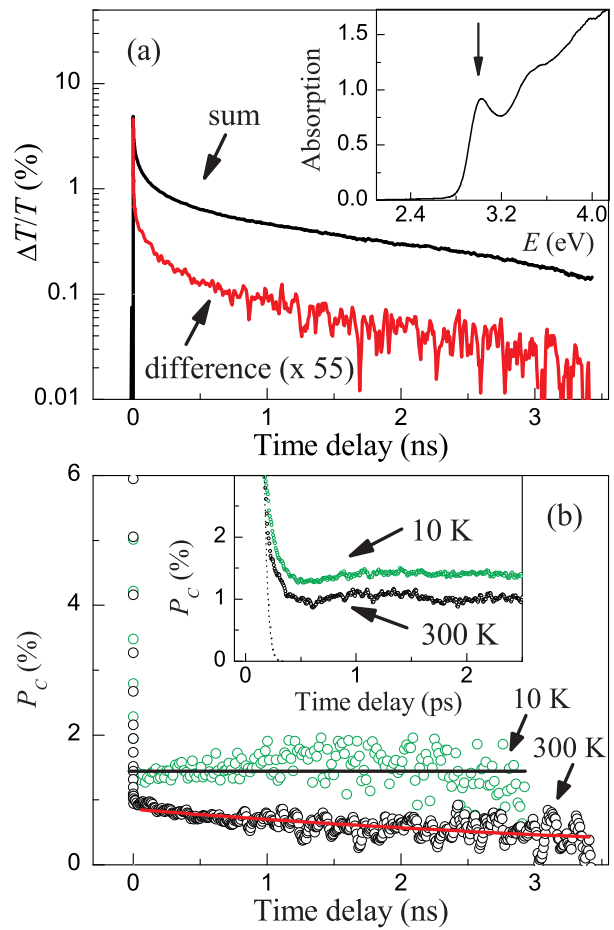}
\caption{ (color online). Spin dynamics in quasi-spherical QDs with typical radius of 2.1~nm. (a) \textit{Sum} and \textit{difference} of differential transmission signals measured at 300~K using co- and counter-circularly polarized probe pulses with respect to the circular polarization of the pump pulses; the \textit{difference} signal was normalized to the initial value of the \textit{sum} signal. The inset shows the optical absorption spectrum of the sample, the arrow indicates the energy, at which the dynamics were measured. (b) Dynamics of {$P_{C}$} at 300~K and 10~K (open points). The curve for 300~K is the exponential decay fit with time constant 5~ns. The horizontal line for 10~K is a guide to the eye. The inset shows the fast initial decay of {the $P_{C}$}, the dotted line illustrates where pump and probe pulses overlap in time (as measured by two-photon absorption). }
\label{fig1}
\end{figure}

In Fig.~1 we show typical room-temperature \textit{difference} and \textit{sum} signals as well as the corresponding dynamics of {the $P_{C}$}. The origin of the spin-insensitive dynamics (\textit{sum} signal) was already investigated in detail~\cite{klimov1999}. Therefore, we will concentrate only on the spin-sensitive signal in the following. However, the rather pronounced decay of the \textit{sum} signal on the investigated time scale [note the logarithmic scale in Fig.~1(a)] implies that correcting for the finite carrier lifetime is essential for an accurate determination of the spin relaxation time. The decay of {the $P_{C}$} on the nanosecond timescale can be well described with an exponential decay function. As seen from the inset in Fig.~1(b), there is also a (sub)picosecond component of {the $P_{C}$} decay beyond the time overlap of pump and probe pulses. As implied by our analysis, this short decay component is not connected to the electron spin relaxation and we treat it in more detail elsewhere~\cite{nahtbp}. Here we concentrate mainly on the long decay component.  The long component of the $P_{C}$ decay at 300~K can be characterized by an exponential decay function with time constant $\tau_{1} \approx$ 5~ns ($\tau_{1}=T_{1}/2$, see Ref.~13
). At lower temperatures the dynamics is slower and also the absolute value of {the $P_{C}$} is larger [see Fig.~1(b)]. In fact, below 200~K the decay of {the $P_{C}$} is so slow that the characteristic decay time cannot be precisely determined within the carrier lifetime. The dependences of the transition rate $\Gamma =1/T_{1}$, corresponding to the time constant $T_{1}$ of the measured $P_{C}$ decay, on the size of the QDs and on the sample temperature are shown in Fig.~2. The spin-relaxation times, inferred from our measurements at temperatures above 200~K, are considerably larger than those in self-assembled QDs~\cite{Tartakovskii04,Gotoh03} and in QDs formed by interface fluctuations in QW~\cite{Stievater02}. Previously reported measurements of spin decoherence in chemically synthesized quasi-spherical QDs also revealed a nanosecond component in the decay of the Faraday rotation~\cite{Gupta01,Gupta02,Stern05} but these measurements were not corrected for the carrier lifetime, which left open questions about the relative importance of the spin-sensitive versus the spin-insensitive contributions to the measured decays. In our undoped samples the absorption of light leads to the generation of neutral excitons. Therefore, the hyperfine interaction with nuclear spins, which is believed to be the dominant spin-relaxation mechanism for electrons and positively charged excitons (trions) in QDs, is ineffective in our case (see, e.g., Ref.~18
).

\begin{figure}[tbp]
\centering \includegraphics[width=8.2cm]{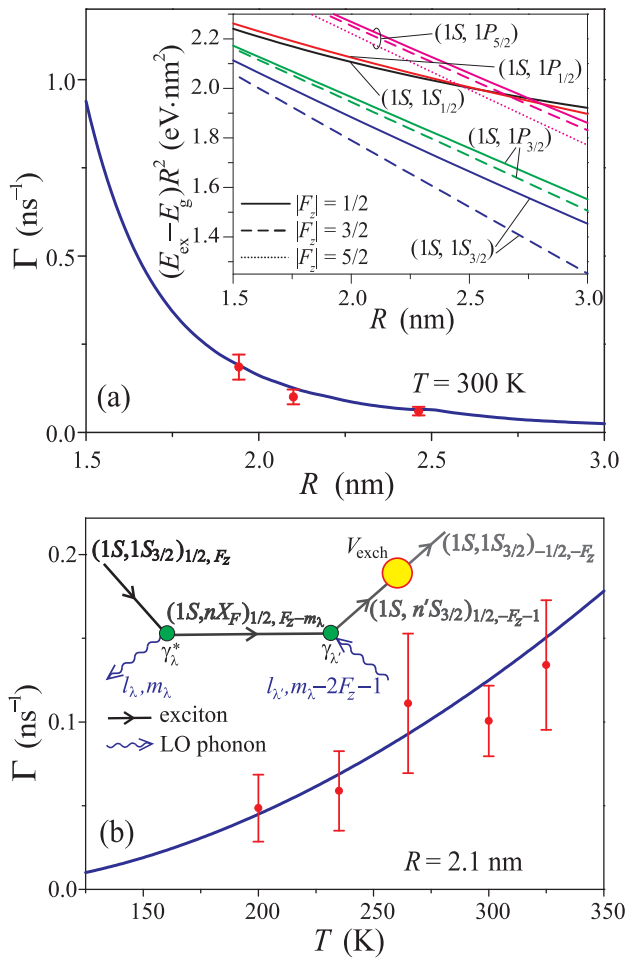}
\caption{(color online). Dependence of the room temperature spin-relaxation rate $\Gamma$ on the QD radius $R$ (a) and temperature dependence of $\Gamma$ in 2.1 nm QDs (b). Points represent the values of $\Gamma =1/T_{1}$ estimated from the measured $P_{C}$ decay and curves show the calculated rate of transitions between the lowest dipole-active exciton states $(1S,1S_{3/2})_{-1/2,3/2}$ and $(1S,1S_{3/2})_{1/2,-3/2}$. {The experimental values of $\Gamma$ and the corresponding error bars were obtained from the processing of multiple data sets measured for each temperature.} Inset to panel (a): size dependence of the exciton energy levels, calculated with the Luttinger parameters $\protect\gamma_1 =1.09$ and $\protect\gamma_2=0.34$~\protect\cite{richard1996} (energies are counted from the band gap of bulk CdS and multiplied by $R^2$).
Inset to panel (b): Feynman diagram {for} the transition $(1S,1S_{3/2})_{1/2,F_z}\to (1S,1S_{3/2})_{-1/2,-F_z}$. The vertices correspond to the electron-hole spin exchange (large circle) and the phonon absorption/emission (small circles). 
The {total} transition amplitude is a sum of contributions, described by six diagrams with different time ordering of the vertices. 
}
\label{fig2}
\end{figure}

\textit{Theoretical model and discussion.} In view of the relatively small spin-orbit splitting in CdS, $\Delta _{\mathrm{so}}\sim 70$~meV \cite{LB}, the six-band model~\cite{grig1990,richard1996} is applied to describe hole states in QDs. A hole, confined to a spherical QD with a cubic lattice, possesses a definite eigenvalue of its total angular momentum $\mathbf{F}$, which is the sum of the hole spin $\mathbf{J}$ ($J=1/2$ for split-off hole bands, $J=3/2$ for light and heavy holes) and the orbital angular momentum of the envelope function $\mathbf{L}$. In the hole wavefunction, components with two different values of the orbital angular momentum, $L$ and $L+2$, are mixed. We label hole states as $nX_{F}$ (cf. Ref.~22,~23
), where the index $X$ refers to the lowest value of the orbital angular momentum in a linear combination, which describes the corresponding state ($X=S,P,D,F,\ldots $ for $ L=0,1,2,3,\ldots $, respectively). The index $n=1,2,3,\ldots $ labels different states with the same $X$ and $F$ in order of increasing energy. While within the four-band model the lowest energy level of a hole in a spherical QD is always $1S_{3/2}$, the six-band model gives for small CdS QDs, with cubic lattice, the lowest energy level $1P_{3/2}$~\cite{grig1990,richard1996}. In QDs with a wurtzite lattice, the crystal field splits the energy levels $nX_{F}$ into sublevels with definite values of $|F_{z}|$~\cite{BP75, efros1996}, where $F_{z}$ is the projection of the total angular momentum of a hole on the $c$-axis of the wurtzite lattice. In bulk CdS the corresponding splitting energy is $\Delta =27$~meV \cite{LB}. For the conduction band, size quantization is relatively strong, and it is sufficient to take into account only the lowest energy level, $1S$, when studying the optical properties of QDs near the absorption edge. Therefore, the relevant states of an electron-hole pair are $(1S,nX_{F})_{s_{z},F_{z}}$, where $s_{z}=\pm 1/2$ is the projection of the electron spin on the $c$-axis. It is important to note that the downward shift of the $(1S,1S_{3/2})_{s_{z},F_{z}}$ states due to the electron-hole Coulomb interaction in a spherical QD~\cite{fomin1998} is significantly larger than that for the $(1S,1P_{3/2})_{s_{z},F_{z}}$ states. As a result, for CdS QDs with $R\gtrsim 1.5$~nm the lowest exciton states turn out to be $(1S,1S_{3/2})_{s_{z},F_{z}}$ [see the inset to Fig.~2(a)].

Only for those dipole-active exciton states, for which the projection of the total angular momentum on the $c$-axis, $N_{z}=F_{z}+s_{z}$, equals $\pm 1$, the optical absorption is influenced by the sense of circular polarization of the absorbed light. For the lowest states of this type, $(1S,1S_{3/2})_{\mp 1/2,\pm 3/2}$, the oscillator strength is 3 times larger than that for the states $(1S,1S_{3/2})_{\pm 1/2,\pm 1/2}$, which are separated from the lowest exciton level by a relatively wide energy spacing (about 23~meV at $R\sim 2$~nm). Therefore, near the absorption edge, the states $(1S,1S_{3/2})_{\mp 1/2,\pm 3/2}$ are significantly more active in the optical absorption compared to higher states with $N_{z}=\pm 1$, so that the dynamics of {the $P_{C}$} is mainly determined by the time evolution of the pump-induced populations of the exciton states $(1S,1S_{3/2})_{-1/2,3/2}$ and $(1S,1S_{3/2})_{1/2,-3/2}$. This population dynamics is strongly affected by the interaction of the excitons with LO phonons. This interaction we describe with the Hamiltonian $H_{\mathrm{int}}=\sum_{\lambda }\left( \gamma_{\lambda }a_{\lambda }+\gamma _{\lambda }^{\ast }a_{\lambda }^{\dagger}\right) $, where $a_{\lambda }^{\dagger }$ and ($a_{\lambda }$) are the creation (annihilation) operators for phonons of the $\lambda $-th mode, and $\gamma _{\lambda }$ are the corresponding interaction amplitudes derived in Ref.~23
. In spherical QDs, both bulk-like and interface phonon modes are characterized by definite values of the quantum numbers $l_{\lambda }$ and $m_{\lambda}$, which are related, respectively, to the phonon angular momentum and its $z$-projection. For CdS quantum dots with $R\sim 2$~nm, the finite lifetime of the LO modes due to their decay into acoustic vibrations causes an LO-mode broadening of about 1~meV~\cite{shiang93}. In general, the mismatch between the exciton-level energy spacings and the LO phonon energy significantly exceeds this LO-mode broadening, so that the probabilities of one-LO-phonon transitions between the lowest exciton states are negligible. In contrast, quasi-elastic two-phonon relaxation processes, in which one LO phonon is absorbed and another is emitted, can lead to fast intralevel relaxation of $F_z$ (without electron spin flip), contributing significantly to the decay of the $P_{C}$ on the (sub)picosecond time scale. For CdS QDs with $R\sim 1.5$ to 3~nm, the room-temperature rates of the two-phonon transitions $(1S,1S_{3/2})_{s_{z},F_{z}}\rightarrow (1S,1S_{3/2})_{s_{z},-F_{z}}$, calculated within the present model, lie in the range 0.2 to 10~ps$^{-1}$. Those values are comparable with the rates of two-phonon transitions between the exciton states $|X\rangle $ and $|Y\rangle$, which dominate the decay of the \textit{linear} polarization of photoluminescence in flat asymmetric InAs/GaAs QDs at temperatures $\gtrsim 100$~K (see Ref.~27). 

However, the $P_C$ dynamics, observed at the {na\-no\-se\-cond} time scale, cannot be explained by the two-phonon transitions discussed above. A key element is that the electron spin polarization, induced by a circularly polarized pump pulse, is not affected by {these transitions}, which eventually lead to a reduction of {the $P_{C}$} to approximately $1/3$ of its initial value{, but not to zero}. We attribute further decay of {the $P_{C}$} to transitions with electron spin flip driven by the electron-hole exchange interaction, which for spherical QDs can be written as~\cite{BP75,efros1996} $V_{\mathrm{exch}}=-\frac{4}{3}\varepsilon _{\mathrm{exch}}a_{o}^{3}\delta \left( \mathbf{r}_{e}-\mathbf{r}_{h}\right) \left( \mathbf{s}\cdot \mathbf{J}\right) $. Here $\mathbf{r}_{e}$ and $\mathbf{s}$ ($\mathbf{r}_{h}$ and $\mathbf{J}$) are the vector coordinates and the spin operator of the electron (hole), $a_{0}$ is the lattice constant of the QD, and $\varepsilon_{\mathrm{exch}}$ is the {\it exchange strength constant}, which can be estimated for CdS as $\varepsilon _{\mathrm{exch}}=35$~meV~\cite{efros1996, broser1980}. Although the exchange interaction is relatively weak (its characteristic energy does not exceed 1~meV in spherical CdS QDs with $R\gtrsim 1.5$~nm), from the present calculations it is found that this interaction provides -- in combination with the exciton-phonon interaction -- an effective channel for exciton spin relaxation processes, which do require electron spin flip. We have identified and calculated the corresponding Feynman diagrams, which give the amplitude of the transition $(1S,1S_{3/2})_{1/2,F_{z}}\rightarrow (1S,1S_{3/2})_{-1/2,-F_{z}}$ {[see inset to Fig.~2(b)]}.

In Fig.~2, we plot the calculated transition rate between the lowest dipole-active exciton states $(1S,1S_{3/2})_{-1/2,3/2}$ and $(1S,1S_{3/2})_{1/2,-3/2}$. The calculations performed take into account transitions, induced by phonon modes with $l_{\lambda },l_{\lambda ^{\prime }}=0,\cdots 5$ (i.e., by $s$-, $p$-, $d$-, $f$-, $g$- and $h$-phonons of a spherical QD -- cf. Ref.~23
), through the intermediate exciton states $(1S,nS_{3/2})$, $(1S,nP_{3/2})$, $(1S,nS_{1/2})$, $(1S,nP_{1/2})$, $(1S,nD_{5/2})$, $(1S,nP_{5/2})$ with $n=1,\cdots 6$. For different sizes of QDs and temperatures, our theoretical results for the transition rate with electron spin flip are in line with the measured $P_{C}$ dynamics. The observed trend towards a reduction of $\Gamma$ with increasing $R$ finds its explanation in the fast (roughly as $R^{-3}$) decrease of the efficiency of the electron-hole exchange interaction with increasing $R$. The temperature dependence is mainly determined by the decreasing number of available LO phonons at lower temperatures. Note that no fitting parameters are used in the calculations. 

\textit{Conclusions.} The dynamics of {the $P_{C}$} in differential transmission, measured in the vicinity of room temperature on quasi-spherical CdS quantum dots, is characterized by two distinct time constants; one on the (sub)picosecond scale, one on the nanosecond scale. We show that the $P_{C}$ decay is governed by two different types of transitions between neutral exciton states. While the \textquotedblleft fast\textquotedblright\ dynamics is determined by processes, where the electron spin is conserved for pump-induced excitons, the \textquotedblleft slow\textquotedblright\ dynamics is attributed in our model to electron spin relaxation. The mechanism of this relaxation suggested in the present paper, namely, two-LO-phonon intralevel transitions with electron spin flip, driven by the electron-hole exchange interaction, is shown to provide a coherent explanation of our experimental results.

This work was supported by the Ministry of Education of the Czech Republic in the framework of the research centre LC510 and the research plan MSM 0021620834, by IUAP, FWO-V projects G.0274.01N, G.0435.03, the WOG WO.035.04N (Belgium), and by the European Commission SANDiE Network of Excellence, contract No. NMP4-CT-2004-500101.

\end{document}